\newcommand{\p}{{\bf p}}
\newcommand{\q}{{\bf q}}
\newcommand{\bp}{{\bf P}}
\newcommand{\Q}{{\bf Q}}

\newcommand{\gdq}{\nabla_{\bf q}}
\newcommand{\gdP}{\nabla_{\bf P}}
\newcommand{\gdQ}{\nabla_{\bf Q}}
\newcommand{\half}{\frac{1}{2}}
\newcommand{\id}{{\bf 1}}
\newcommand{\lvb}{\Bigl \bracevert}

\documentstyle[12pt, fullpage]{article}

\begin{document}

\title{A PDE-Based Approach to Classical Phase-Space Deformations}

\author{Emmanuel Tannenbaum\\
{\it Harvard University, Cambridge, MA 02138}}

\maketitle

\begin{abstract}

This paper presents a PDE-based approach to finding an optimal
canonical basis with which to represent a nearly integrable
Hamiltonian.  The idea behind the method is to continuously deform the
initial canonical basis in such a way that the dependence of the Hamiltonian 
on the canonical position of the final basis is minimized.  The
final basis incorporates as much of the classical dynamics as possible
into an integrable Hamiltonian, leaving a much smaller non-integrable
component than in the initial representation.  With this approach it
is also possible to construct the semiclassical wavefunctions
corresponding to the final canonical basis.  This optimized basis is
potentially useful in quantum calculations, both as a way to minimize
the required size of basis sets, and as a way to provide physical
insight by isolating those effects resulting from integrable dynamics.

\end{abstract}

\section{Introduction}

Suppose we are given a nearly integrable Hamiltonian $ H(\q, \p) $,
where $ (\q, \p) $ represents some canonical representation of phase
space (not necessarily ordinary position and momentum).  We wish to
find another canonical representation $ (\Q, \bp) $ in which $ H $ is
as close as possible to being integrable.  Our motivation for this is
twofold, and is connected to semiclassical quantum mechanics: First of
all, from a purely numerical perspective, a representation in which $
H $ is as close as possible to being integrable leads to an optimized
semiclassical basis with which to perform quantum calculations.  The
reason for this is that we can associate with $ \bp $ a quantum state
$ |\bp\rangle $.  If we write $ H(\Q, \bp) = H^{(0)}(\bp) +
H^{(1)}(\Q, \bp) $, then it is clear that $ H^{(0)} $ is diagonal in
the $ \{|\bp\rangle\} $ basis, so that only $ H^{(1)} $ is available
to couple the various basis states.  The coupling is given
semiclassically by \cite{key8},
\begin{equation}
\langle \bp'|\hat{H}^{(1)}|\bp \rangle = H^{(1)}_{\frac{\bp'-\bp}{2\pi\hbar}}
(\frac{\bp+\bp'}{2})
\end{equation}
where $ H^{(1)}_{\frac{\bp'-\bp}{2\pi\hbar}}(\frac{\bp+\bp'}{2}) $ is
the $ \frac{\bp'-\bp}{2\pi\hbar} $ Fourier component at $
\frac{\bp+\bp'}{2} $ of $ H^{(1)} $.  A representation in which $
H^{(1)} $ is as small as possible will minimize the couplings in the
corresponding $ \{|\bp\rangle\} $ basis, and thus will minimize the
size of the basis required to perform a given calculation to some
desired accuracy.

Secondly, a canonical represenation in which $ H $ is as close as
possible to being integrable provides physical insight.  By
incorporating as much of the classical dynamics as possible into an
integrable Hamiltonian, this optimal canonical representation can help
to isolate those classical and quantum effects resulting from
integrable dynamics from those that do not.  Thus this representation
can isolate non-integrable quantum and classical effects such as
dynamical tunneling and Arnol'd diffusion, respectively.  Furthermore,
we can also visualize the quantum manifestation of the integrable
classical dynamics via the semiclassical prescription for constructing
a wavefunction: Given the generating function $ S(\q, \bp) $ from the
initial basis $ (\q, \p) $ to the final basis $ (\Q, \bp) $, we have,
up to normalization,
\begin{equation}
\langle \q| \bp\rangle = |\mbox{det} \frac{\partial^2 S}
{\partial \q \partial \bp}|^{\half}\exp[iS/\hbar]
\end{equation}
where $ \mbox{det} \frac{\partial^2 S}{\partial \q \partial \bp} $ is the
well-known Van Vleck determinant.

In the case of action-angle variables, the optimized representation of
a nearly integrable Hamiltonian is termed an Intrinsic Resonance
Representation (IRR), a term coined by Carioli, Heller, and Moller
(CHM) \cite{key1, key2}.  In 1997 they published a paper detailing an
algorithm for the construction of such a representation.  The idea
behind the CHM algorithm is to eliminate all the non-resonant terms of
the Hamiltonian via an appropriate canonical transformation.  This
canonical transformation is obtained via a modified Chapman, Garrett,
and Miller (CGM) method \cite{key1, key2, key3}, which is essentially a
Newton-Raphson scheme to find the invariant tori with a desired set of
actions for a nearly integrable system.  The remaining resonant and
near-resonant terms are then re-expressed in the new basis.  It is
impossible to reduce the angle dependence any further, since this
would result in the formation of resonance zones, which prevents a
global action-angle description of the Hamiltonian.

In 2001 the author published an alternative algorithm for finding the
IRR basis \cite{key1}.  The method introduced is a PDE-based approach
which continuously deforms the initial action-angle basis in such a
way that the angle dependence of the Hamiltonian is continuously
reduced.  It amounts to a gradient-descent algorithm in the limit of a
first-order perturbation, and was therefore called the GDA method.
Formally, the method does not distinguish between resonant, nearly
resonant, and non-resonant terms, that is, the evolution is performed
on the entire Hamiltonian without any terms neglected.  However, the
evolution is such that the more non-resonant a term, the more strongly
it is affected by the evolution.  Thus, the non-resonant terms of the
Hamiltonian are essentially killed off, the nearly resonant terms are
reduced somewhat, while the resonant terms are essentially unaffected.
The GDA method was successfully applied in Ref. 1 to 2, 4, and 6
degree-of-freedom systems.

The GDA method circumvents two main drawbacks of the CHM method.
First, the CHM method requires an a priori decision as to which terms
are resonant and non-resonant.  This leads to an ambiguity in the case
of near-resonances.  It could happen that a given term in the
Hamiltonian must be considered resonant in order to get the
Newton-Raphson scheme to converge.  This leads to a somewhat
artificial cutoff criterion, since a nearly resonant Fourier component
should in principle still be reduced as much as possible, though not
necessarily completely.  Thus, unless the Hamiltonian has a few exact
or near-resonances, it is not clear that the CHM method will give the
optimized torus basis.

Second, the CHM method requires the numerical evaluation of
multidimensional integrals, and the numerical inversion of a nonlinear
angle map at every iteration step.  These numerical calculations slow
the algorithm down.  In contrast, the numerical calculations required
by the GDA method are much simpler, so we believe that the GDA method 
is faster than the CHM approach (though in fairness it should be
added that no direct speed comparisons have been made to date).

Despite its advantages, the GDA method is limited to systems describable
by action-angle variables.  Furthermore, while the method generates
the representation of the Hamiltonian in the final action-angle basis,
it does not give the overall generating function $ S(\q, \bp) $,
transforming from the initial $ (\q, \p) $ basis to the final $ (\Q,
\bp) $ basis.  Thus, we could compute the energy spectrum arising from
an optimized invariant torus basis, but not the corresponding
semiclassical wavefunctions ``living'' on the IRR tori.

This paper generalizes the GDA method to arbitrary canonical
representations of phase space, and extends it to include the
determination of the overall generating function $ S(\q, \bp) $.  It
is organized as follows: In Section 2 we derive the generic evolution
equation for the Hamiltonian starting from an arbitrary canonical
basis $ (\q, \p) $.  We then go on to derive the corresponding
evolution equation for the overall generating function, with which it
is possible to construct semiclassical wavefunctions.  In Section 3 we
consider the case of a nearly integrable Hamiltonian, and obtain the
specific form our evolution is to take if we want to optimize the
canonical basis used to represent the Hamiltonian.  In Section 4 we
consider the first-order limit of our PDE approach.  In particular, we
obtain a generalized first-order classical perturbation theory which coincides
with classical perturbation theory in the case of action-angle variables.  In
Section 5 we consider some analytical examples.  We look at free
propagation in one dimension perturbed by a localized potential, and
also in two dimensions perturbed by a symmetric Gaussian bump.  We
continue in Section 6 with a numerical example.  Finally, we conclude
in Section 7 with a summary of our results and a discussion of future
research plans.

\section{The Evolution Equations}

In this section we shall derive the basic evolution equations for the
Hamiltonian $ H $ and the overall generating function $ S $.  The idea
is as follows: We start with an initial set of canonical coordinates $
(\q, \p) $, which denote any global representation of phase space, and
do not necessarily refer to ordinary position and momentum,
respectively (non-global representations can also work, as long as we
remain well within the region of phase space they describe.  An
example of this is action-angle variables for a system which can
dissociate.  Action-angle variables should work as a valid
representation as long as we remain well below any dissociation
threshold).  We continuously deform this system via a series of
infinitesimal generating functions.  The result is that our canonical
representation is evolving with time, and is denoted by $ (\Q_t,
\bp_t) $ at time $ t $.  As our canonical pair evolves, the functional
dependence of $ H $ on the canonical pair changes.  In addition, the
overall generating function $ S(\q, \bp; t) $ connecting the initial $
(\q, \p) $ to the current $ (\Q_t, \bp_t) $ evolves as well.  In
subsection 2.1, we derive the PDE governing the evolution of $ H $
generated by this phase space deformation.  In subsection 2.2, 
we derive the PDE governing the evolution of $ S $.

\subsection{Evolution of the Hamiltonian}

Consider an arbitrary set of canonical coordinates.  At time $ t $, we're at 
system  $ (\Q_t, \bp_t) $.  At time $ t + dt $, we're at system $ (\Q_{t+dt},
\bp_{t+dt}) $.  These are connected by an infinitesimal generating function 
$ F(\Q_t, \bp_{t+dt}) = \Q_t \cdot \bp_{t+dt} + dt G(\Q_t, \bp_{t+dt})
$.  Therefore,
\begin{equation}
\Q_{t+dt} = \Q_t + dt \gdP G(\Q_t, \bp_{t+dt})
\end{equation}
\begin{equation}
\bp_t = \bp_{t+dt} + dt \gdQ G(\Q_t, \bp_{t+dt})
\end{equation}
so to first-order, we obtain,
\begin{equation}
\Q_t = \Q_{t+dt} - dt \gdP G(\Q_{t+dt}, \bp_{t+dt})
\end{equation}
\begin{equation}
\bp_t = \bp_{t+dt} + dt \gdQ G(\Q_{t+dt}, \bp_{t+dt})
\end{equation}
The Hamiltonian at time $ t + dt $ is therefore related to the
Hamiltonian at time $ t $ via,
\begin{eqnarray}
H^{(t+dt)}(\Q_{t+dt}, \bp_{t+dt}) & = & H^{(t)}(\Q_t, \bp_t) \nonumber \\
                    		  & = & H^{(t)}(\Q_{t+dt} - dt \gdP 
                                                G(\Q_{t+dt}, \bp_{t+dt}),
 	    		 		        \bp_{t+dt} + dt \gdQ 
                                                G(\Q_{t+dt}, \bp_{t+dt})) 
					\nonumber \\
                    		  & = & H^{(t)}(\Q_{t+dt}, \bp_{t+dt}) + dt
                                        (\gdP H^{(t)} \cdot \gdQ G -
					 \gdQ H^{(t)} \cdot \gdP G)
\end{eqnarray}
and so,
\begin{equation}
\frac{\partial H}{\partial t} = \gdP H \cdot \gdQ G - \gdQ H \cdot \gdP G 
                              = -\{H, G\}
\end{equation}
which can be re-written as,
\begin{equation}
\frac{\partial H}{\partial t} + \{H, G\} = 0
\end{equation}
Note that the evolution equation relates the value of $ H^{(t+dt)} $ at
$ (\Q_{t+dt}, \bp_{t+dt}) $ to the value of $ H^{(t)} $ at $ (\Q_{t+dt},
\bp_{t+dt}) $.  Thus, this evolution generates a one-parameter family 
$ \{H(\Q, \bp; t)\} $ of Hamiltonians, whose evolution under the
action of the infinitesimal generating functions $ G(\Q, \bp; t) $ is
given by the previous equation (Eq. (9)).

For what follows in this paper it will prove convenient to represent
the dynamics in Fourier space.  To this end, assume that $ H $ is
periodic in each $ Q_i $ with period $ L_i $.  Then we shall choose $
G $ to also be periodic in each $ Q_i $ with period $ L_i $.  Define $
V = L_1 \cdots L_D $, and let $ \Omega $ denote an arbitrary
D-dimensional box of side lengths $ L_1, \dots, L_D $.  Then,
\begin{equation}
H(\Q, \bp) = \frac{1}{V}\sum_{\bf k}{H_{\bf k}(\bp)e^{2\pi i {\bf k} \cdot \Q}}
\end{equation}
where
\begin{equation}
H_{\bf k}(\bp) \equiv \int_{\Omega}{d\Q H(\Q, \bp)e^{-2\pi i {\bf k} \cdot \Q}}
\end{equation}
and similarly for $ G(\Q, \bp) $.

The full evolution done component-wise gives,
\begin{eqnarray}
\frac{1}{V}\sum_{\bf k}{\frac{\partial H_{\bf k}}{\partial t}
e^{2\pi i {\bf k} \cdot \Q}} & = & \frac{1}{V^2}(\sum_{\bf k'}{\gdP H_{\bf k'}
e^{2\pi i {\bf k'} \cdot \Q}} \cdot 2\pi i \sum_{\bf k''}{{\bf k''}G_{\bf k''}
e^{2\pi i {\bf k''} \cdot \Q}} - \nonumber \\
&   & 2\pi i \sum_{\bf k'}{{\bf k'}H_{\bf k'}e^{2\pi i {\bf k'} \cdot \Q}} 
\cdot \sum_{\bf k''}{\gdP G_{\bf k''} e^{2\pi i {\bf k''} \cdot \Q}})
\end{eqnarray}
which gives,
\begin{eqnarray}
\sum_{\bf k}{\frac{\partial H_{\bf k}}{\partial t}e^{2\pi i {\bf k} \cdot \Q}}
& = & 
\frac{2\pi i}{V}\sum_{{\bf k'}, {\bf k''}}{[({\bf k''} \cdot \gdP H_{\bf k'})
G_{\bf k''} - ({\bf k'} \cdot \gdP G_{\bf k''})H_{\bf k'}]
e^{2\pi i ({\bf k'} + {\bf k''}) \cdot \Q}} \nonumber \\
& = &
\frac{2\pi i}{V}\sum_{\bf k}{\sum_{\bf k'}{[({\bf k'} \cdot \gdP H_{{\bf k} -
{\bf k'}})G_{\bf k'} - (({\bf k} - {\bf k'}) \cdot \gdP G_{\bf k'})H_{{\bf k} -
{\bf k'}}]e^{2\pi i {\bf k} \cdot \Q}}}
\end{eqnarray}
Our component-wise evolution is therefore,
\begin{equation}
\frac{\partial H_{\bf k}}{\partial t} = 2\pi i \frac{1}{V}
\sum_{\bf k'}{[({\bf k'} \cdot \gdP H_{{\bf k} - {\bf k'}})G_{\bf k'} -
               (({\bf k} - {\bf k'}) \cdot \gdP G_{\bf k'})
	       H_{{\bf k} - {\bf k'}}]} 
\end{equation}
Note that the evolution preserves the integration limits of our
original system.  Thus the topology of the original phase space is
preserved.

As a final note for this subsection, it should be mentioned that
degrees of freedom for which $ L_i $ is finite can be treated in an
action-angle formalism in which $ L_i = 1 $, while degrees of freedom
for which $ L_i = \infty $ have their corresponding Fourier sums
replaced by integrals.

\subsection{Evolution of the Generating Function}

The above treatment tells us how the Hamiltonian evolves under the
action of the infinitesimal generating functions, but it tells us
nothing of the evolution of $ S({\bf q}, {\bf P}; t) $, the overall
generating function from the initial system $ (\q, \p) $ to $ (\Q_t,
\bp_t) $.  We shall deal with this now.

At time $ t $, our generating function is $ S(\q, \bp; t) $, taking us
from $ (\q, \p) $ to $ (\Q_t, \bp_t) $.  At time $ t + dt $, our
generating function is $ S(\q, \bp; t+dt) $, taking us from $ (\q, \p)
$ to $ (\Q_{t+dt}, \bp_{t+dt}) $.  We know that $ (\Q_t, \bp_t) $ and
$ (\Q_{t+dt}, \bp_{t+dt}) $ are connected by an infinitesimal
generating function $ \Q_t \cdot
\bp_{t+dt} + dt G(\Q_t, \bp_{t+dt}) $.  We must have,
\begin{equation}
\p = \frac{\partial S}{\partial \q}(\q, \bp_t; t) = 
     \frac{\partial S}{\partial \q}(\q, \bp_{t+dt}; t+dt)
\end{equation}

Now, 
\begin{eqnarray}
\frac{\partial S}{\partial \q}(\q, \bp_t; t) & = & \frac{\partial S}
{\partial \q}(\q, \bp_{t+dt} + dt\frac{\partial G}{\partial \Q}(\Q_t,
\bp_{t+dt}); t) \nonumber \\ 
					     & = & \frac{\partial S}
{\partial \q}(\q, \bp_{t+dt}; t) + \frac{\partial^2 S}{\partial \bp 
\partial \q}(\q, \bp_{t+dt}; t) \cdot \frac{\partial G}{\partial \Q}
(\Q_t, \bp_{t+dt}) dt \nonumber \\
                                             & = & \frac{\partial S}
{\partial \q}(\q, \bp_{t+dt}; t) + \frac{\partial G}{\partial \Q}
(\Q_t, \bp_{t+dt}) \cdot \frac{\partial^2 S}{\partial \q \partial \bp}
(\q, \bp_{t+dt}; t) dt \nonumber \\
                                             & = & \frac{\partial S}
{\partial \q}(\q, \bp_{t+dt}; t) + \frac{\partial G}{\partial \Q}
(\Q(\q, \bp_t; t), \bp_{t+dt}) \cdot \frac{\partial \Q}{\partial \q}
(\q, \bp_{t+dt}; t) dt \nonumber \\
                                             & = & \frac{\partial S}
{\partial \q}(\q, \bp_{t+dt}; t) + \frac{\partial G}{\partial \Q}
(\Q(\q, \bp_{t+dt}; t), \bp_{t+dt}) \cdot \frac{\partial \Q}{\partial \q}
(\q, \bp_{t+dt}; t) dt \nonumber \\
                                             & = & \frac{\partial S} 
{\partial \q}(\q, \bp_{t+dt}; t) + \frac{\partial G}{\partial \q}
(\Q(\q, \bp_{t+dt}; t), \bp_{t+dt}) dt
\end{eqnarray}
The transformation between the second and third lines in the above derivation
was obtained by switching from column vectors to row vectors.

We also have,
\begin{equation}
\frac{\partial S}{\partial \q}(\q, \bp_{t+dt}; t+dt) = \frac{\partial S}
{\partial \q}(\q, \bp_{t+dt}; t) + \frac{\partial^2 S}{\partial t \partial \q}
(\q, \bp_{t+dt}; t) dt
\end{equation}
and so,
\begin{equation}
\frac{\partial^2 S}{\partial t \partial \q} = \frac{\partial G}{\partial \q}
(\Q(\q, \bp_{t+dt}; t), \bp_{t+dt})
\end{equation}

We also have,  
\begin{eqnarray}
\Q_t & = & \frac{\partial S}{\partial \bp}(\q, \bp_t; t) \nonumber \\
     & = & \frac{\partial S}{\partial \bp}(\q, \bp_{t+dt} + dt \frac
           {\partial G}{\partial \Q}(\Q_t, \bp_{t+dt}); t) \nonumber \\
     & = & \frac{\partial S}{\partial \bp}(\q, \bp_{t+dt}; t) +
           \frac{\partial^2 S}{\partial \bp^2}(\q, \bp_{t+dt}; t)
           \cdot \frac{\partial G}{\partial \Q}(\Q_t, \bp_{t+dt}) dt
\end{eqnarray} 
and
\begin{eqnarray} 
\Q_{t+dt} & = & \frac{\partial S}{\partial \bp}(\q, \bp_{t+dt}; t+dt) 
                \nonumber \\
          & = & \frac{\partial S}{\partial \bp}(\q, \bp_{t+dt}; t) + 
		\frac{\partial^2 S}{\partial t \partial \bp}
                (\q, \bp_{t+dt}; t) dt
\end{eqnarray}
and 
\begin{eqnarray}
\Q_{t+dt} & = & \Q_t + \frac{\partial G}{\partial \bp}(\Q_t, \bp_{t+dt}) dt
                \nonumber \\
          & = & \frac{\partial S}{\partial \bp}(\q, \bp_{t+dt}; t) +
                \frac{\partial^2 S}{\partial \bp^2}(\q, \bp_{t+dt}; t)
                \cdot \frac{\partial G}{\partial \Q}(\Q_t, \bp_{t+dt}) dt +
                \frac{\partial G}{\partial \bp}(\Q_t, \bp_{t+dt}) dt
\end{eqnarray}
and so,
\begin{eqnarray}
\frac{\partial^2 S}{\partial t \partial \bp}(\q, \bp_{t+dt}; t) & = &
\frac{\partial^2 S}{\partial \bp^2}(\q, \bp_{t+dt}; t) \cdot 
\frac{\partial G}{\partial \Q}(\Q_t, \bp_{t+dt}) +
\frac{\partial G}{\partial \bp}(\Q_t, \bp_{t+dt}) \nonumber \\
                                                                 & = &
\frac{\partial G}{\partial \Q}(\Q(\q, \bp_t; t), \bp_{t+dt}) \cdot
\frac{\partial \Q}{\partial \bp}(\q, \bp_{t+dt}; t) +
\frac{\partial G}{\partial \bp}(\Q(\q, \bp_t; t), \bp_{t+dt}) \nonumber \\
                                                                 & = &
\frac{\partial G}{\partial \Q}(\Q(\q, \bp_{t+dt}; t), \bp_{t+dt}) \cdot
\frac{\partial \Q}{\partial \bp}(\q, \bp_{t+dt}; t) +
\frac{\partial G}{\partial \bp}(\Q(\q, \bp_{t+dt}; t), \bp_{t+dt}) \nonumber \\
                                                                 & = &
\frac{\partial G}{\partial \bp}(\Q(\q, \bp_{t+dt}; t), \bp_{t+dt})\lvb _{\q}
\end{eqnarray}
Once again, a switch from column to row vectors gives us the equality between
the first and second lines of the above derivation.  Therefore, we see that,
\begin{equation}
\frac{\partial^2 S}{\partial \q \partial t} = \frac{\partial G}{\partial \q}
(\frac{\partial S}{\partial \bp}(\q, \bp; t), \bp)
\end{equation}
and
\begin{equation}
\frac{\partial^2 S}{\partial \bp \partial t} = \frac{\partial G}{\partial \bp}
(\frac{\partial S}{\partial \bp}(\q, \bp; t), \bp)\lvb _{\q}
\end{equation}
so we obtain,
\begin{equation}
\frac{\partial S}{\partial t}(\q, \bp; t) = G(\frac{\partial S}{\partial \bp}
(\q, \bp; t), \bp) + C(t)
\end{equation}
Since we only care about $ (\Q_t, \bp_t) $ and not about $ S $ itself, we can
drop the $ C(t) $ term, giving us, finally,
\begin{equation}
\frac{\partial S}{\partial t} = G(\frac{\partial S}{\partial \bp}, \bp)
\end{equation}

At time $ t = 0 $, $ S $ is simply the identity transformation, so
that $ S(\q, \bp; 0) = \q \cdot \bp $.  Thus, $ S $ is not periodic in
each $ q_i $ with period $ L_i $.  Define $ {\bf n}{\bf L} = (n_1 L_1,
\dots, n_D L_D) $, and note, however, that $ S(\q + {\bf n}{\bf L},
\bp; 0) = \bp \cdot {\bf n} {\bf L} + S(\q + {\bf n}{\bf L}, \bp; 0)
$.  We claim that this property is preserved by the evolution.  To see
this, suppose that at time $ t $ we have $ S(\q + {\bf n}{\bf L}, \bp;
t) = \bp \cdot {\bf n}{\bf L} + S(\q, \bp; t) $.  Then,
\begin{equation}
\frac{\partial S}{\partial \bp}(\q + {\bf n}{\bf L}, \bp; t) = 
{\bf n}{\bf L} + \frac{\partial S}{\partial \bp}(\q, \bp; t)
\end{equation}
The periodicity of $ G $ then implies,
\begin{equation}
\frac{\partial S}{\partial t}(\q + {\bf n}{\bf L}, \bp; t) = 
\frac{\partial S}{\partial t}(\q, \bp; t)
\end{equation}
Therefore, since $ S(\q + {\bf n}{\bf L}, \bp; 0) = \bp \cdot {\bf
n}{\bf L} + S(\q, \bp; 0) $, it follows from the previous equation
that this property holds at all $ t $.  Thus, although $ S $ is not
periodic in the $ q_i $'s, we still need only track $ S $ for $ \q $
in a D-dimensional box of side lengths $ L_1, \dots, L_D $.

\section{Choosing $ G $}

We want an approach that minimizes the dependence of $ H $ on $ \Q $.
The condition that $ H $ be independent of $ \Q $ is equivalent to the
vanishing of the nonzero Fourier components $ H_{\bf k}(\bp) $.  We
therefore seek to minimize the dependence of $ H $ on $ \Q $ by
choosing $ G $ in such a way that the $ |H_{\bf k}(\bp; t)| $ are
continuously decreasing for all $ {\bf k} \neq 0 $.

At some time $ t $, we can write $ H(\Q, \bp; t) = H^{(0)}(\bp; t) +
H^{(1)} (\Q, \bp; t) $.  The idea is that $ H^{(0)} $ contains the
piece of the Hamiltonian which is only dependent on $ \bp $, and $
H^{(1)} $ contains the remainder.  Then,
\begin{eqnarray}
\frac{\partial H}{\partial t} & = & (\gdP H^{(0)} + \gdP H^{(1)}) \cdot
				    \gdQ G - \gdQ H^{(1)} \cdot \gdP G
                                    \nonumber \\
                              & = & \gdP H^{(0)} \cdot \gdQ G +
                                    \gdP H^{(1)} \cdot \gdQ G -
                                    \gdQ H^{(1)} \cdot \gdP G \nonumber \\
                              & = & \gdP H^{(0)} \cdot \gdQ G -
                                    \{H^{(1)}, G\}
\end{eqnarray}

In the limit of a first-order perturbation on an integrable Hamiltonian, the 
relevant equation is,
\begin{equation}
\frac{\partial H}{\partial t} = \gdP H^{(0)} \cdot \gdQ G
\end{equation}
In Fourier space, this becomes,
\begin{equation}
\frac{\partial H}{\partial t} = \frac{1}{V}\sum_{\bf k}{\frac{
\partial H_{\bf k}}{\partial t} e^{2\pi i {\bf k} \cdot \Q}} =
\gdP H^{(0)} \cdot \frac{2\pi i}{V} \sum_{\bf k}{{\bf k} G_{\bf k}
e^{2\pi i {\bf k} \cdot \Q}}
\end{equation}
so
\begin{equation}
\frac{\partial H_{\bf k}}{\partial t} = 2\pi i ({\bf k} \cdot \gdP H^{(0)})
G_{\bf k}
\end{equation}
Then $ \frac{\partial H_{\bf k}\bar{H}_{\bf k}}{\partial t} =
H_{\bf k} \frac{\partial \bar{H}_{\bf k}}{\partial t} +
\bar{H}_{\bf k} \frac{\partial H_{\bf k}}{\partial t} =
2\pi i ({\bf k} \cdot \gdP H^{(0)})(\bar{H}_{\bf k} G_{\bf k} - H_{\bf k}
\bar{G}_{\bf k}) $.  Therefore, in the first-order limit, the gradient-descent
prescription for minimizing $ |H_{\bf k}|^2 = H_{\bf k}\bar{H}_{\bf
k}, {\bf k} \neq 0 $ is to set $ G_{\bf k} = 2\pi i ({\bf k} \cdot
\gdP H^{(0)}) H_{\bf k} $.  Then $ \bar{G}_{\bf k} = -2\pi i ({\bf k}
\cdot \gdP H^{(0)}) \bar{H}_{\bf k} $, so in the first-order limit we
obtain that,
\begin{equation}
\frac{\partial H_{\bf k}\bar{H}_{\bf k}}{\partial t} = -8\pi^2({\bf k} \cdot
\gdP H^{(0)})^2 H_{\bf k}\bar{H}_{\bf k}
\end{equation}
which is clearly negative.  For stronger perturbations, this is no longer the
gradient-descent prescription.  However, for nearly integrable systems (the
ones of interest to us in this paper) the perturbation should still be 
sufficiently weak that the above choice for $ G_{\bf k} $ will shrink the 
$ H_{\bf k}\bar{H}_{\bf k}, {\bf k} \neq 0 $.  Therefore, we take,
\begin{equation}
G_{\bf k} = 2\pi i ({\bf k} \cdot \gdP H^{(0)}) H_{\bf k}
\end{equation}
This gives,
\begin{eqnarray}
G(\Q, \bp) & = & \frac{1}{V}\sum_{\bf k}{G_{\bf k}(\bp) e^{2\pi i {\bf k} 
                 \cdot \Q}} 
             =   \frac{1}{V} 2\pi i \gdP H^{(0)} \cdot \sum_{\bf k}
                 {{\bf k}H_{\bf k}e^{2\pi i {\bf k} \cdot \Q}} \nonumber \\
           & = & \frac{2\pi i}{V} \gdP H^{(0)} \cdot \frac{1}{2\pi i}\gdQ
                 \sum_{\bf k}{H_{\bf k}e^{2\pi i {\bf k} \cdot \Q}} 
             =   \gdP H^{(0)} \cdot \gdQ H 
\end{eqnarray}
so $ G(\Q, \bp) = \gdP H^{(0)} \cdot \gdQ H = \gdP H^{(0)} \cdot \gdQ
H^{(1)} $.  Note that the Fourier expansion of $ G $ involves terms of
the form $ {\bf k} \cdot \gdP H^{(0)} $.  A $ {\bf k} $ for which $
{\bf k} \cdot \gdP H^{(0)} = 0 $ is a generalized resonance at $ \bp
$.  The integrability of $ H^{(0)} $ is destroyed by the resonant
terms in $ H^{(1)} $.

Our evolution does not formally distinguish between resonances,
near-resonances, and non-resonances.  The evolution is done on the
entire Hamiltonian without any terms neglected.  However, the closer a
term is to being resonant, the smaller the corresponding Fourier
component of $ G $, and so the less that term is affected by the
evolution.

In the first-order limit, $ H^{(0)}(\bp; t) $ differs from $
H^{(0)}(\bp; 0) $ by a correction which is at most first-order in $
H^{(1)} $.  Therefore, if we use $ H^{(0)}(\bp; 0) $ instead of $
H^{(0)}(\bp; t) $ in our prescription for choosing $ G $ in Eq. (35),
we get a discrepancy of at most second-order in $ H^{(1)} $, so that
the two formulations are equivalent to first-order.  Since our
prescription for choosing $ G $ was derived from the first-order limit
of the evolution of $ H $, we see that it is equivalent to use $
H^{(0)} (\bp; 0) $ or $ H^{(0)}(\bp; t) $ in Eq. (35).  Finally, our $
t = 0 $ Hamiltonian is usually given as $ H(\q, \p) = H_0(\p) + V(\q,
\p) $, where $ H_0 $ is the zeroth-order, integrable Hamiltonian, and
$ V $ is the perturbation.  We can extract the $ {\bf k} = 0 $ Fourier
component of $ V $, writing $ V(\q, \p) = V_0(\p) + \tilde{V}(\q, \p)
$.  Then $ H(\Q, \bp; 0) = H_0(\bp) + V_0(\bp) + \tilde{V}(\Q, \bp) $,
so that $ H^{(0)}(\bp; 0) = H_0(\bp) + V_0(\bp) $, and $ H^{(1)}(\Q,
\bp) = \tilde{V}(\Q, \bp) $.  Therefore, note that in the first-order
limit it is equivalent to use $ H^{(0)}(\bp; 0) $ or $ H_0(\bp) $ in
the prescription for choosing $ G $.  Once again, this means that it
is equivalent to use $ H^{(0)}(\bp; 0) $ or $ H_0(\bp) $.  In what
follows $ H^{(0)}(\bp; t) $, $ H^{(0)}(\bp; 0) $, and $ H_0(\bp) $
will all be denoted by $ H^{(0)} $, or $ H^{(0)}(\bp) $.  When
required, we will specify to which $ H^{(0)} $ we are referring.

We conclude this section by deriving the PDE governing the evolution
of $ S $, given our prescription for choosing $ G $.  We have,
\begin{equation}
\frac{\partial S}{\partial t}(\q, \bp; t) = G(\Q(\q, \bp; t), \bp; t) =
\gdP H^{(0)} \cdot \gdQ H(\Q(\q, \bp; t), \bp; t) 
\end{equation}
Now, we know that $ H(\Q, \bp; t) = H(\q, \p; 0) = H(\q, \frac{\partial S}{
\partial \q}(\q, \bp; t); 0) $.  We also have, 
\begin{equation} 
\frac{\partial}{\partial \q} = \frac{\partial}{\partial \Q} \cdot 
\frac{\partial \Q}{\partial \q} = \frac{\partial}{\partial \Q} \frac
{\partial^2 S}{\partial \q \partial \bp}
\end{equation}
Switching from row to column vectors gives $ \frac{\partial}{\partial \q} =
\frac{\partial^2 S}{\partial \bp \partial \q}\frac{\partial}{\partial \Q} $,
and so,
\begin{equation}
\frac{\partial}{\partial \Q} = (\frac{\partial^2 S}{\partial \bp \partial \q})
^{-1} \frac{\partial}{\partial \q}
\end{equation}
and so the dynamics of $ S $ is governed by,
\begin{equation}
\frac{\partial S}{\partial t} = \frac{\partial H^{(0)}}{\partial \bp} \cdot 
(\frac{\partial^2 S}{\partial \bp \partial \q})^{-1} \frac{\partial H(\q,
\frac{\partial S}{\partial \q}; 0)}{\partial \q}\lvb _{\bp}
\end{equation}
Now, $ \frac{\partial H(\q, \frac{\partial S}{\partial \q}; 0)}
{\partial \q}\lvb _{\bp} = \frac{\partial H}{\partial \q}\lvb _{\p}(\q, \frac{\partial
S}{\partial \q}; 0) + \frac{\partial H}{\partial \p}\lvb _{\q}(\q, \frac{\partial 
S}{\partial \q}; 0) \cdot \frac{\partial^2 S}{\partial \q^2} $.  Once again,
switching from row to column vectors gives,
\begin{equation}
\frac{\partial H(\q, \frac{\partial S}{\partial \q}; 0)}{\partial \q}\lvb _{\bp} =
\frac{\partial H}{\partial \q}\lvb _{\p}(\q, \frac{\partial S}{\partial \q}; 0) + 
\frac{\partial^2 S}{\partial \q^2} \frac{\partial H}{\partial \p}\lvb _{\q}
(\q, \frac{\partial S}{\partial \q}; 0) 
\end{equation}
Note that since $ H(\q, \p; 0) $ is simply our initial Hamiltonian, this PDE
for $ S $ involves $ S $ only.  We do not need the evolution of $ H $ in order
to get the evolution of $ S $.

The numerical evolution of $ S $ is described in Appendix A.  The
numerical evolution of $ H $ in the case of action-angle variables is
discussed at length in Ref. 1.  Since the case for arbitrary canonical
pairs is handled similarly, we do not give numerical details for the $
H $ evolution in this paper.

\section{The First-Order Limit}

From our choice of $ G $ in the previous section, it follows that in the 
limit of a first-order perturbation on an integrable Hamiltonian,
\begin{equation}
\frac{\partial H_{\bf k}}{\partial t} = 
-4\pi^2 ({\bf k} \cdot \gdP H^{(0)})^2 H_{\bf k}
\end{equation}
Our first-order solution yields $ H_{\bf k}(\bp; t) = H_{\bf k}(\bp; 0) 
\exp[-4\pi^2 ({\bf k} \cdot \gdP H^{(0)})^2 t] $.  Note that the more
non-resonant a term, the faster the exponential decay.  In particular,
resonances are not affected at all.

We now turn to the evolution of $ S $ in the first-order limit.  To
this end, write $ S(\q, \bp; t) = \q \cdot \bp + G(\q, \bp; t) $.  In 
what follows we shall work to first-order in $ G $ and $ H^{(1)} $.
Then,
\begin{equation}
\frac{\partial H}{\partial \q}\lvb _{\p}(\q, \frac{\partial S}{\partial \q}; 0) = 
\frac{\partial H}{\partial \q}(\q, \bp; 0) + \frac{\partial^2 H}{\partial \bp
\partial \q}(\q, \bp; 0) \frac{\partial G}{\partial \q}
\end{equation}
and
\begin{eqnarray}
\frac{\partial^2 S}{\partial \q^2}\frac{\partial H}{\partial \p}\lvb _{\q}
(\q, \frac{\partial S}{\partial \q}; 0) = 
\frac{\partial^2 G}{\partial \q^2}\frac{\partial H}{\partial \p}\lvb _{\q}
(\q, \bp + \frac{\partial G}{\partial \q}; 0) =
\frac{\partial^2 G}{\partial \q^2}\frac{\partial H}{\partial \bp}(\q, \bp; 0)
\end{eqnarray}
Now, $ H(\q, \p; 0) = H^{(0)}(\p; 0) + H^{(1)}(\q, \p; 0) $.  Then $
\frac{\partial H}{\partial \q}(\q, \bp; 0) = \frac{\partial H^{(1)}}
{\partial \q}(\q, \bp; 0) $.  To first-order, $ \frac{\partial^2 H}
{\partial \bp \partial \q}(\q, \bp; 0) \frac{\partial G}{\partial \q}
=
\frac{\partial^2 H^{(1)}}{\partial \bp \partial \q}(\q, \bp; 0) \frac
{\partial G}{\partial \q} = 0 $.  Finally, to first-order, $
\frac{\partial^2 G}{\partial \q^2}\frac{\partial H}{\partial \bp}\lvb
_{\q}(\q, \bp; 0) =
\frac{\partial^2 G}{\partial \q^2}\frac{\partial H^{(0)}}{\partial \bp}
(\bp; 0) $.  Since each of these terms are either first-order in $
H^{(1)} $ or $ G $, in the first-order limit we take $
(\frac{\partial^2 S}{\partial
\bp \partial \q})^{-1} = \id $.  Putting everything together gives us that our
first-order equation is,
\begin{equation}
\frac{\partial G}{\partial t} = \frac{\partial H^{(0)}}{\partial \bp} 
\cdot (\frac{\partial H^{(1)}}{\partial \q} + \frac{\partial^2 G}
{\partial \q^2}\frac{\partial H^{(0)}}{\partial \bp})
\end{equation}
In Fourier space, this becomes,
\begin{equation}
\frac{\partial G_{\bf k}}{\partial t} =
2\pi i ({\bf k} \cdot \gdP H^{(0)}) H^{(1)}_{\bf k} - 4\pi^2 ({\bf k} \cdot 
\gdP H^{(0)})^2 G_{\bf k}
\end{equation}
Since $ G_{\bf k}(\bp; 0) = 0 \mbox{ }\forall \mbox{ }{\bf k} $, we
obtain,
\begin{equation}
G_{\bf k}(\bp; t) = \frac{i H^{(1)}_{\bf k}}{2\pi ({\bf k} \cdot \gdP 
H^{(0)})}(1 - e^{-4\pi^2 ({\bf k} \cdot \gdP H^{(0)})^2 t})
\end{equation}
Note that $ G_{\bf k}(\bp; t) = 0 $ for all resonant terms.  This can
be seen by looking at the original ODE from which the solution is
derived, or equivalently by noting that $ \lim_{{\bf k} \cdot \gdP
H^{(0)} \rightarrow 0} $ $ {(1 - \exp[-4\pi^2 ({\bf k} \cdot \gdP
H^{(0)})^2 t])/({\bf k} \cdot \gdP H^{(0)})} = 0 $.  We can write,
\begin{equation}
G(\q, \bp; t) = \frac{1}{V}\frac{i}{2\pi}\sum_{{\bf k} \neq 0}{\frac{H^{(1)}
_{\bf k}}{{\bf k} \cdot \gdP H^{(0)}}(1 - e^{-4\pi^2 ({\bf k} \cdot \gdP 
H^{(0)})^2t})e^{2\pi i{\bf k} \cdot \q}}
\end{equation}
This series is convergent, because the exponential term prevents
resonances and near resonances in the denominator from causing the
series to diverge.  In practice, in the first-order limit we only run
$ t $ out to some finite value, so that the Fourier components of $ G
$ remain small.  This smallness criterion depends on what is the
maximum allowable $ G $ before a first-order approximation is no
longer deemed accurate, and is therefore dependent on the specific
error cutoffs for the system at hand.  The more non-resonant terms in
the Hamiltonian will essentially get killed off, while the less
non-resonant terms will get reduced somewhat, though not completely.
Exact resonances will be unaffected by the evolution.  In any event we
can let $ t \rightarrow \infty $ to get the first-order perturbation 
theory result,
\begin{equation}
G(\q, \bp) = \frac{1}{V}\frac{i}{2\pi}\sum_{{\bf k} \neq {\bf 0}}
{\frac{H^{(1)}_{\bf k}}{{\bf k} \cdot \gdP H^{(0)}} e^{2\pi i {\bf k} 
\cdot \q}}
\end{equation}
where the sum is over all non-resonant $ {\bf k} $.  Note, however,
that the convergence of the various Fourier components to their $ t 
\rightarrow \infty $ limits is not uniform, because the time constant
for the exponential term is proportional to $ 1/({\bf k} \cdot
\gdP H^{(0)})^2 $.  This goes to infinity as $ {\bf k} $ approaches
a resonance.  

In reality, the above equation must be solved for finite $ t $, and
then take the $ t \rightarrow \infty $ limit.  This prevents any kind
of ambiguities in $ G $, something which will be illustrated with one
of the analytical examples in the next section.

Before concluding this section, we should note the similarity between
this generalized first-order perturbation theory and classical
first-order perturbation theory in the special case of action-angle
variables.  Indeed, our generalized first-order perturbation theory
reduces to classical first-order perturbation theory in the case of
action-angle variables.  One wonders if there is a corresponding
KAM-like Theorem for more general ``tori'' than just those described
by action-angle variables.

\section{Analytical Examples}

\subsection{1-D Example}

Consider the case of free propagation perturbed by a weak, localized
potential.  Our Hamiltonian is,
\begin{equation}
H(q, p) = \frac{p^2}{2m} + V(q)
\end{equation}
where $ V(q) $ is finite in $ {\bf R} $ and decays to $ 0 $ as $ q \rightarrow
\pm \infty $.  For above-barrier energies we can construct an action function
defined everywhere in coordinate space, given by,
\begin{equation}
S(q, P) = P\int^{q}{(1 - \frac{2mV(q')}{P^2})^\half dq'}
\end{equation}
where $ P \equiv p(\pm \infty) $.  Along a classical trajectory, we
have $ \frac{p^2}{2m} + V(q) = H(q, p) = \frac{p_{\infty}^2}{2m} =
\frac{P^2}{2m} $, so the representation of $ H $ in the corresponding
$ (Q, P) $ system is simply $ H(q, p) = H^{(0)}(P) = \frac{P^2}{2m} $.

Because $ S(q, P) $ is only determined up to some (possibly
P-dependent) constant term, all we can uniquely specify is $ p =
\frac{\partial S}{\partial q} = P(1 - \frac{2mV(q)}{P^2})^\half $.  In
the first-order limit, we know from the previous section that our PDE
method converges to a steady-state.  Our method should still come close
to a steady-state beyond the first-order, yet still weakly perturbed,
regime.  For the strongly perturbed case, we don't know what our
method will do.  Nevertheless, we can show that for all above-barrier
energies, our method has a unique steady-state solution, and is given
by the formula for $ p(q, P) $ given above.  To prove this, we note
that in one dimension the evolution equation for $ S $ becomes,
\begin{equation}
\frac{\partial S}{\partial t} = \frac{P}{m}\frac{1}{\frac{\partial^2 S}
{\partial P \partial q}}\frac{\partial H}{\partial q}\lvb _P
\end{equation}
Setting $ \frac{\partial S}{\partial t} = 0 $, and remembering that
for energies above the barrier we have $ P \neq 0 $, we get the
steady-state equation,
\begin{equation}
\frac{\partial H(q, \frac{\partial S}{\partial q})}{\partial q}\lvb _P = 0 
\end{equation}
Note that this assumes that $ \frac{\partial^2 S}{\partial P \partial
q} \neq 0, \infty $, something that we will only be able to check once
we solve our equation.  The steady-state equation can be integrated at
constant $ P $ to give $ H(q, \frac{\partial S}{\partial q}) = C(P) $.
Letting $ q \rightarrow
\pm \infty $ gives us that $ H $ becomes $ \frac{p_{\infty}^2}{2m} = \frac{P^2}
{2m} $, so $ C(P) = \frac{P^2}{2m} $.  Therefore, $ \frac{1}{2m}(\frac{\partial
S}{\partial q})^2 + V(q) = \frac{P^2}{2m} \Rightarrow \frac{\partial S}
{\partial q} = P(1 - \frac{2mV(q)}{P^2})^{\half} $.  

Note that $ \frac{\partial^2 S}{\partial P \partial q} \neq 0, \infty
$ for all energies above the barrier, so our PDE approach does yield
the appropriate steady-state solution.  Below the barrier, we cannot
construct a real action function defined for all $ q $, since the
momentum $ p $ becomes imaginary in the classically forbidden region.
At a turning point, $ \frac{\partial^2 S} {\partial P \partial q} =
\infty $, so in any event our method for finding the steady-state
breaks down below the barrier.  Thus, both approaches to finding a
real action function $ S $ presented in this section have an equal
range of validity, namely, for all energies above the barrier.
Semiclassically, this means that our PDE approach recovers WKB theory
for above-barrier energies.  Maitra and Heller \cite{key10} developed
a method to compute above-barrier reflection coefficients using the
WKB wavefunctions as a distorted-wave basis.  The above derivation
shows that this approach is contained as a subcase within our 
PDE-based approach.

\subsection{2-D Example}

Consider the Hamiltonian,
\begin{equation}
H(x, y, p_x, p_y) = \frac{p_x^2 + p_y^2}{2m} + \lambda e^{-\alpha(x^2 + y^2)}
\end{equation}
We wish to construct an action function $ S(x, y, P_x, P_y) $ in the
first-order limit.  By the symmetry of this problem, we need only
consider $ P_y = 0 $.  So, let's write $ S(x, y, P_x, P_y) = xP_x +
yP_y + G(x, y, P_x, P_y) $, where $ G $ is our first-order correction.
We substitute into the time-independent Hamilton-Jacobi equation (HJE)
to get,
\begin{eqnarray}
E & = & H(x, y, P_x + \frac{\partial G}{\partial x}, 
                P_y + \frac{\partial G}{\partial y}) \nonumber \\
  & = & \frac{1}{2m}(P_x^2 + P_y^2) + 
        \frac{P_x}{m}\frac{\partial G}{\partial x} + 
        \frac{P_y}{m}\frac{\partial G}{\partial y} + 
        \lambda e^{-\alpha(x^2 + y^2)}
\end{eqnarray} 
Set $ P_y = 0 $, $ E = \frac{P_x^2}{2m} $ to get,
\begin{equation}
\frac{\partial G}{\partial x} = -\frac{\lambda m}{P_x} e^{-\alpha(x^2 + y^2)}
\end{equation}
which can be integrated to give $ G(x, y, P_x, 0) = -\frac{\lambda
m}{P_x} e^{-\alpha y^2}\int_{-\infty}^{x}{e^{-\alpha x'^2} dx'} + C(y,
P_x) $.  Now, $ C $ is determined by the boundary conditions which we
impose on this system.  The first set of boundary conditions we will
consider is the requirement that $ p_y = 0 $ at $ x = -\infty $.  We
have,
\begin{equation}
\frac{\partial G}{\partial y} = \frac{2\lambda m \alpha}{P_x} y e^{-\alpha y^2}
\int_{-\infty}^{x}{e^{-\alpha x'^2} dx'} + \frac{\partial C}{\partial y}
\end{equation}
Then at $ x = -\infty $ we get $ p_y = \frac{\partial C}{\partial y} =
0 $, so $ C = C(P_x) $.  Since the HJE only depends on the momentum
field generated by $ S $ and not on $ S $ itself, a $ P_x $-dependent
constant term is unimportant, so we can set it to $ 0 $.  Therefore,
one first-order solution is,
\begin{eqnarray}
G_1(x, y, P_x, 0) & = & -\frac{\lambda m}{P_x}e^{-\alpha y^2}\int_{-\infty}^{x}
                        {e^{-\alpha x'^2} dx'} \nonumber \\
                  & = & -\frac{\lambda m}{P_x \sqrt{\alpha}}e^{-\alpha y^2}
 			\int_{-\infty}^{\sqrt{\alpha} x}{e^{-u^2} du}
\end{eqnarray}

An alternative solution is obtained from the boundary condition that $
p_y\lvb _{x = -\infty} = -p_y\lvb _{x = \infty} $.  Plugging into our
expression for $ \frac{\partial G}{\partial y} $ gives $
\frac{\partial C} {\partial y} = -\frac{2\lambda m
\alpha}{P_x}\sqrt{\frac{\pi}{\alpha}} y e^{-\alpha y^2} -
\frac{\partial C}{\partial y} $, which can be solved for $
\frac{\partial C}{\partial y} $ and integrated to yield,
\begin{equation}
C(y, P_x) = \frac{\lambda m}{2 P_x}\sqrt{\frac{\pi}{\alpha}}e^{-\alpha y^2} +
            \tilde{C}(P_x)
\end{equation}
As before, we drop the $ P_x $-dependent constant term.  The result is
another first-order solution,
\begin{eqnarray}
G_2(x, y, P_x, 0) & = & -\frac{\lambda m}{P_x \sqrt{\alpha}}e^{-\alpha y^2}
                        (\int_{-\infty}^{\sqrt{\alpha} x}{e^{-u^2} du} -
		         \frac{\sqrt{\pi}}{2}) \nonumber \\
                  & = & -\frac{\lambda m}{P_x \sqrt{\alpha}}e^{-\alpha y^2}
                        (\int_{-\infty}^{\sqrt{\alpha} x}{e^{-u^2} du} -
			 \int_{-\infty}^{0}{e^{-u^2} du}) \nonumber \\
                  & = & -\frac{\lambda m}{P_x \sqrt{\alpha}}e^{-\alpha y^2}
			\int_{0}^{\sqrt{\alpha} x}{e^{-u^2} du}
\end{eqnarray}

Figures 1a,b, plot the momentum fields generated by $ G_1 $ and $ G_2
$.  Note that $ G_1 $ produces the more physically intuitive manifold
of trajectories, since they first head straight toward the Gaussian
bump, and are only deflected as they approach it.  The $ G_2 $
trajectories, in contrast, are symmetrical to the left and right of
the bump.  This is accomplished by initially angling the trajectories
inward toward the bump.  The trajectories curve in and then curve away
as they reach the bump.

We now turn to a first-order treatment using the approach derived from
our PDE method.  We begin by using the position formulation of our PDE
in the first-order limit, given by Eq. (44).  We know that the
evolution of $ G $ goes to a steady-state, so setting $ \frac{\partial
G}{\partial t} = 0 $, and plugging in the terms for our specific
potential, yields,
\begin{equation}
-\frac{2 \alpha \lambda}{m}e^{-\alpha(x^2 + y^2)}(P_x x + P_y y) +
\frac{1}{m^2}(\frac{\partial^2 G}{\partial x^2}P_x^2 + 
              2\frac{\partial^2 G}{\partial x \partial y} P_x P_y +
              \frac{\partial^2 G}{\partial y^2}P_y^2) = 0
\end{equation}
At $ P_y = 0 $ we get $ \frac{\partial^2 G}{\partial x^2} = \frac{2 \alpha 
\lambda m}{P_x} x e^{-\alpha(x^2 + y^2)}$.  Integrating gives,
\begin{equation}
\frac{\partial G}{\partial x} = -\frac{\lambda m}{P_x} e^{-\alpha(x^2 + y^2)} +
                                C(y, P_x)
\end{equation}
If we require $ p_x\lvb _{x = -\infty} = P_x $, then $ \frac{\partial
G} {\partial x}\lvb _{x = -\infty} = 0 \Rightarrow C(y, P_x) = 0 $.
Our resulting differential equation for $ G $ is therefore identical
to the one obtained via the HJE in the first-order limit.

We now use the Fourier formulation of our first-order, PDE-based
approach, by applying Eq. (47).  The Fourier components of $
e^{-\alpha (x^2+y^2)} $ are given by $
\lambda\frac{\pi}{\alpha}e^{-\frac{\pi^2}{\alpha}(k_x^2 + k_y^2)} $.
Therefore, we obtain,
\begin{equation}
G(x, y, P_x, P_y; t) = \frac{i\lambda}{2\pi}\frac{\pi}{\alpha}\int_{-\infty}^
	               {\infty}{\int_{-\infty}^{\infty}{
		       {dk_x dk_y \frac{\exp[-\frac{\pi^2}{\alpha}
                       (k_x^2 + k_y^2)]}{k_x\frac{P_x}{m} + 
                       k_y\frac{P_y}{m}}(1 - e^{-4\pi^2\frac{(k_x P_x + 
                        k_y P_y)^2}{m^2}t})e^{2\pi i(k_x x + k_y y)}}}}
\end{equation}
We now set $ P_y = 0 $, giving
\begin{eqnarray}
G(x, y, P_x, 0; t) & = & \frac{i\lambda m}{2\alpha P_x}\int_{-\infty}^{\infty}
                         {\int_{-\infty}^{\infty}
			 {dk_x dk_y\frac{\exp[-\frac{\pi^2}{\alpha}k_x^2]
                         \exp[-\frac{\pi^2}{\alpha}k_y^2]}{k_x}
                         (1 - e^{-4\pi^2 k_x^2 P_x^2 t/m^2})}}
                         e^{2\pi i k_x x}e^{2\pi i k_y y} \nonumber \\
                   & = & \frac{i\lambda m}{2\alpha P_x}\sqrt{\frac{\alpha}
                         {\pi}}e^{-\alpha y^2}\int_{-\infty}^{\infty}
	                 {dk_x \frac{\exp[-\frac
                         {\pi^2}{\alpha}k_x^2]}{k_x}(1 - e^{-4\pi^2 k_x^2 
                         P_x^2 t/m^2})e^{2\pi i k_x x}} \nonumber \\
                   & = & \frac{i\lambda m}{2 P_x \sqrt{\alpha \pi}}
                         e^{-\alpha y^2}\int_{-\infty}^{\infty}
                         {dk_x e^{-\frac{\pi^2}{\alpha} 
                         k_x^2}(1 - e^{-4\pi^2 k_x^2 P_x^2 t/m^2})
                         (2\pi i \int_{0}^{x}{e^{2\pi i k_x x'} dx'} + 
                         \frac{1}{k_x})} \nonumber \\
                   & = & -\frac{\lambda m}{P_x}\sqrt{\frac{\pi}{\alpha}}
		         e^{-\alpha y^2}\int_{0}^{x}{dx'\int_{-\infty}^{\infty}
                         {dk_x e^{-\frac{\pi^2}{\alpha}k_x^2}(1 - 
                         e^{-4\pi^2 k_x^2 P_x^2 t/m^2})e^{2\pi i k_x x'}}} + 
                         \nonumber \\
                   &   & \frac{i\lambda m}{2 P_x \sqrt{\alpha \pi}}
		         e^{-\alpha y^2}\int_{-\infty}^{\infty}
                         {\frac{dk_x}{k_x} e^{-\frac{\pi^2}
                         {\alpha} k_x^2}(1 - e^{-4\pi^2 k_x^2 P_x^2 t/m^2})}
\end{eqnarray} 
The second integral vanishes, because the integrand is an odd function of
$ k_x $.  This gives us,
\begin{eqnarray}
G(x, y, P_x, 0; t) & = & -\frac{\lambda m}{P_x}\sqrt{\frac{\pi}{\alpha}}
	                 e^{-\alpha y^2}\int_{0}^{x}{dx'\int_{-\infty}^
                         \infty}{dk_x e^
		         {-\frac{\pi^2}{\alpha}k_x^2}(1 - e^{-4\pi^2 k_x^2 
                         P_x^2 t/m^2})e^{2\pi i k_x x'}} \nonumber \\
                   & = & -\frac{\lambda m}{P_x}\sqrt{\frac{\pi}{\alpha}}
                         e^{-\alpha y^2}\int_{0}^{x}{dx'\int_{-\infty}^{\infty}
                         {dk_x (\exp[-\frac{\pi^2}{\alpha}k_x^2] - 
                         \exp[-\frac{\pi^2}{\frac{1}{\frac{1}{\alpha} +
                         4P_x^2 t/m^2}}k_x^2])e^{2\pi i k_x x'}}} \nonumber \\
                   & = & -\frac{\lambda m}{P_x}\sqrt{\frac{\pi}{\alpha}}
                         e^{-\alpha y^2}\int_{0}^{x}dx' 
                         (\sqrt{\frac{\alpha}{\pi}}\exp[-\alpha x'^2] - 
                         \nonumber \\
                   &   & \frac{1}{\sqrt{\pi(\frac{1}{\alpha} + 4P_x^2 t/m^2)}}
                         \exp[-\frac{x'^2}{\frac{1}{\alpha} + 4P_x^2 t/m^2}])
\end{eqnarray}

We can break this integral into two pieces.  First let's note the
following change of variable: Setting $ u = \sqrt{\alpha}\lambda $
gives, $ \int_{0}^{x}{d\lambda \sqrt{\frac{\alpha}{\pi}}e^{-\alpha
\lambda^2}} =
\frac{1}{\sqrt{\pi}}\int_{0}^{\sqrt{\alpha}x}{du e^{-u^2}} $.  Therefore,
\begin{eqnarray}
G(x, y, P_x, 0; t) & = & -\frac{\lambda m}{P_x\sqrt{\alpha}}e^{-\alpha y^2}
		         (\int_{0}^{\sqrt{\alpha}x}{e^{-u^2}du} -
                          \int_{0}^{\frac{x}{\sqrt{\frac{1}{\alpha} + 
                          4 P_x^2t/m^2}}}{e^{-u^2}du}) \nonumber \\
                   & = & -\frac{\lambda m}{P_x\sqrt{\alpha}}e^{-\alpha y^2}
                         \int_{\frac{\sqrt{\alpha}x}{\sqrt{1 + 
                         4 P_x^2 \alpha t/m^2}}}^{\sqrt{\alpha}x}{e^{-u^2}du}
\end{eqnarray}
Letting $ t \rightarrow \infty $ gives us,
\begin{equation}
G(x, y, P_x, 0) = -\frac{\lambda m}{P_x\sqrt{\alpha}}e^{-\alpha y^2}
                   \int_{0}^{\sqrt{\alpha}x}{e^{-u^2}du} 
\end{equation}
Thus, the Fourier space approach yields the $ G_2 $ function obtained
previously.  Note that it was necessary to take the $ t \rightarrow \infty $
limit only at the end, in order to avoid any ambiguities in evaluating the
integrals.

\section{A Numerical Example}

We chose to test the $ S $ evolution numerically on the
two-dimensional Pullen-Edmonds Hamiltonian \cite{key7}, given by,
\begin{equation}
H(x, y, p_x, p_y) = \frac{p_x^2 + p_y^2}{2} + \half(\omega_x^2 x^2 + 
\omega_y^2 y^2) + \epsilon x^2 y^2 
\end{equation}
This was the two-dimensional system studied numerically in Ref. 1 in
testing the GDA method.  The results there were compared with those
obtained in Ref. 2 using the CHM method.  However, while in Ref. 1 the
evolution was done on $ H $ in the context of obtaining a
semiclassical quantum spectrum, in this case it is the PDE for $ S $
that is being tested.

We set $ \omega_x = \omega_y = 1 $.  The $ t = 0 $ canonical representation is
simply the harmonic oscillator action-angle basis, denoted by $ (\theta_x,
\theta_y, J_x, J_y) $.  The arbitrary $ t $ canonical representation is denoted
by $ (\phi_x, \phi_y, I_x, I_y) $, so that $ S = S(\theta_x, \theta_y,
I_x, I_y; t) $, with $ J_x = \frac{\partial S}{\partial \theta_x} $,
and $ J_y =
\frac{\partial S}{\partial \theta_y} $.  The transformation to the harmonic 
representation is obtained by setting $ x = \sqrt{\frac{J_x}{\pi}}\cos 2\pi 
\theta_x $, $ p_x = -\sqrt{\frac{J_x}{\pi}}\sin 2\pi \theta_x $, and similarly
for $ y, p_y $.  The result is,
\begin{equation}
H(\theta_x, \theta_y, J_x, J_y) = \frac{1}{2\pi}(J_x + J_y) + 
\frac{\epsilon J_x J_y}{4\pi^2}(1 + \cos 4\pi \theta_x + \cos 4\pi \theta_y +
\half(\cos 4\pi(\theta_x + \theta_y) + \cos 4\pi (\theta_x - \theta_y)))
\end{equation}
The only nonvanishing Fourier components are $ \pm (2,0), \pm (0,2),
\pm (2,2) $, and $ \pm (2,-2) $.  Note in particular that $ \pm (2,-2)
$ is a resonance.

Of the three prescriptions for choosing $ G $, the simplest one to
use is $ H^{(0)} = H_0 = \frac{1}{2\pi}(1,1) \cdot (I_x, I_y) $, so
that $ \gdP H^{(0)} $ is taken to be $ \frac{1}{2\pi}(1,1) $.  In
addition, for this Hamiltonian it is readily verified that,
\begin{eqnarray}
\frac{\partial H}{\partial \q}\lvb _{\p}(\q, \frac{\partial S}{\partial \q}; 0)
& = & -\frac{\epsilon}{\pi}
\frac{\partial S}{\partial \theta_x}\frac{\partial S}{\partial \theta_y}
(\sin 4\pi \theta_x + \half(\sin 4\pi(\theta_x + \theta_y)
+ \sin 4\pi(\theta_x - \theta_y)), \nonumber \\
&   & 
 \sin 4\pi \theta_y + \half(\sin 4\pi(\theta_x + \theta_y) 
- \sin 4\pi(\theta_x - \theta_y)))
\end{eqnarray}
and,
\begin{equation}
\frac{\partial H}{\partial \p}\lvb _{\q} = \frac{1}{2\pi}(1,1) + 
\frac{\epsilon}
{4\pi^2}(1 + \cos 4\pi \theta_x + \cos 4\pi \theta_y + \half(\cos 4\pi
(\theta_x + \theta_y) + \cos 4\pi (\theta_x - \theta_y)))
(\frac{\partial S}{\partial \theta_y}, \frac{\partial S}{\partial
\theta_x})
\end{equation}
We substitute into Eq. (40) and then into Eq. (39) to get the PDE for
$ S $.

We set $ \epsilon = 0.05 $, and $ I_x = I_y = 9\pi $.  This
corresponds to a torus with zeroth-order energy $ E = 9 $ at $ t = 0
$.  The system is nearly integrable in this regime \cite{key2}, yet a
study of the semiclassically obtained energy spectrum \cite{key1,
key2} shows clear differences with first-order perturbation theory.
We therefore test our PDE beyond the first-order limit with this
example.

We set $ N = 20 $, $ DX = 0.1 $, $ DT = 0.034 $, and $ GDSZ = 9 $,
which allowed us to propagate the PDE out to a time of $ 0.306 $ (the
meaning of these parameters is given in Appendix A).  The rate of
change of $ S $ on the grid reaches a minimum at this point (which was
determined by tracking $ \sqrt{\langle (\partial S/\partial t)^2\rangle} $
on the grid), so the evolution was stopped here.  Because our PDE approach
is only the gradient-descent prescription in the first-order limit, for
finite perturbations there is no reason to expect the evolution to reach
steady-state.  We discuss this issue further in Appendix C.

Figures 2a-d show the results of the evolution at times $ t = 0.0,
0.1, 0.2 $, and $ 0.3 $.  It should be noted that we are not plotting
$ S $ in these graphs.  Rather, we plot $ H(\theta_x, \theta_y,
\frac{\partial S}{\partial \theta_x}, \frac{\partial S} {\partial
\theta_y}) $.  There are two reasons for this:  First, because we are working in
the weakly perturbed, though still beyond first-order, regime, $ S $
remains fairly close to the identity transformation.  The effect on $
H(\theta_x, \theta_y, \frac{\partial S}{\partial \theta_x},
\frac{\partial S} {\partial \theta_y}) $, however, is dramatic, and so
it is much more convenient to represent the evolution of $ S $ in this
indirect fashion.  Second, even if the perturbation were sufficiently
strong to significantly deform $ S $, the only way to determine if the
deformation of $ S $ is correct is to look at its effect on $ H $.

Using the formula derived in Appendix B, it is readily shown that the
first-order solution to $ H(\theta_x, \theta_y, \frac{\partial
S}{\partial \theta_x},
\frac{\partial S}{\partial \theta_y}) $ is,
\begin{eqnarray}
H(\theta_x, \theta_y, \frac{\partial S}{\partial \theta_x},
\frac{\partial S}{\partial \theta_y}) & = &
\frac{1}{2\pi}(I_x + I_y) + \frac{\epsilon I_x I_y}{4\pi^2}
(1 + e^{-4t}(\cos 4\pi \theta_x + \cos 4\pi \theta_y) + \nonumber \\
                                      &   &
\half(\cos 4\pi (\theta_x - \theta_y) + e^{-16t} 
\cos 4\pi (\theta_x + \theta_y)))
\end{eqnarray}
While the perturbation is sufficiently strong that there are clear
differences with first-order perturbation theory, the perturbation is
still sufficiently weak that the first-order result provides a
qualitative and semiquantitative picture of how the evolution should proceed.  
Letting $ t \rightarrow \infty $, we see that the long-time limit of the 
evolution gives,
\begin{equation}
H(\theta_x, \theta_y, \frac{\partial S}{\partial \theta_x},
\frac{\partial S}{\partial \theta_y})\lvb _{t = \infty} =
\frac{1}{2\pi}(I_x + I_y) + \frac{\epsilon I_x I_y}{4\pi^2}(1 +
\half \cos 4\pi (\theta_x - \theta_y))
\end{equation}
This is plotted for our specific set of parameters in Figure 3.  Note
that the numerical evolution does indeed deform $ S $ in such a way
that the graphs in Figures 2a-d evolve to look like the graph in
Figure 3.  Without the $ (2, -2) $ resonance, H is integrable at this
energy \cite{key2}, so it is possible to transform to a basis in which
$ H $ only depends on $ (I_x, I_y) $.  Instead, the evolution
eliminates as much of the nonresonant behavior as possible, but the
resonant angle dependence arising from the $ (2, -2) $ term remains.
In Figures 4a-d we present the results of the evolution on the
Hamiltonian obtained by removing the $ (2, -2) $ resonance from the
Pullen-Edmonds term.  We changed $ DT $ to $ 0.033 $ and $ GDSZ $ to $
11 $.  All other parameters are otherwise unchanged.  In this case,
the evolution should flatten $ H $, and as Figures 4a-d confirm, this
is exactly what happens.

\section{Conclusions and Future Research}

This paper presented a PDE-based, phase-space deformation approach to
optimize the canonical basis with which to globally represent a nearly
integrable Hamiltonian.  This paper is a generalization of the GDA
method in that we are no longer restricted to action-angle variables.
Any canonical representation of phase space can be used, making the
GDA approach adaptable to Hamiltonians other than just those
describable in an action-angle formalism.  This paper also extends the
GDA method because it now provides an expression for the evolution of
$ S $, the overall generating function connecting the initial
canonical basis to the final canonical basis.  Because semiclassical
wavefunctions are constructed from classical action functions via
Eq. (2), we now have an approach for constructing the wavefunctions
associated with the optimized torus basis.

The purpose of this paper was to develop the PDE-based approach in
complete generality, and then to demonstrate it via analytical results
in the first-order regime and a simple numerical example in the
nearly integrable regime.  The emphasis in this paper was on the PDE
for the evolution of $ S $, because the PDE for $ H $ was tested
extensively in Ref. 1 in the case of action-angle variables.

The PDEs for $ H $ and $ S $ involve quantities that are purely
classical.  Thus, there was no quantum mechanics in this paper.
Nevertheless, as was mentioned previously, the motivation for this
work is derived from semiclassical quantum mechanics.  Ref. 1 has
already used the $ H $ evolution as a method to optimize semiclassical
torus bases for use in quantum calculations.  For future work, we seek
to use the $ S $ evolution as a tool to actually construct the
semiclassical wavefunctions.  We also hope to apply our approach to
real systems.  One area where this PDE-based approach might be useful
is in the determination of vibrational energies of polyatomic
molecules.

The generality of the PDE-based approach means that other
previous methods are contained as subcases.  As mentioned in Section
5.1, this approach contains as a subcase the use of one-dimensional
WKB wavefunctions as a distorted-wave basis for computing
above-barrier reflection coefficients.  This method was developed by
Maitra and Heller in Ref. 10.  In their paper, they raised the issue
of generalizing their technique to higher dimensions, and to
action-angle systems.  Our PDE-based approach is exactly this
generalization, since it unifies the Maitra and Heller method and the
GDA method under one general approach.  Furthermore, the PDE-based
approach also reduces to classical first-order perturbation theory in
the limit of a first-order perturbation on an integrable Hamiltonian.

To conclude, we should add that Heller has often made the comparison
between the separatrix region generated by a local potential bump in
one dimension to the resonance zone structure in a Poincare surface of
section of a nearly integrable Hamiltonian \cite{key6, key10}.
Indeed, Heller regards the above-barrier reflection problem as a
prototype for the more complicated case of dynamical tunneling between
invariant tori facilitated by resonance zones.  Because both types of
systems can be treated within the same PDE-based approach, they are,
in fact, formally equivalent phenomena.

\section{Acknowldegments}

This research was supported by the Department of Chemistry and
Chemical Biology at Harvard, the Department of Physics at Harvard, and
by an NSF Graduate Research Fellowship.  The author would like to
thank E.J. Heller for providing the inspiration for this work, and for
helpful discussions leading to its completion. The author would also
like to thank Prof. Anthony Yezzi, Jr.  (of the Georgia Tech
Department of Electrical and Computer Engineering) for helpful
conversations regarding the numerical solution of the PDEs.

\appendix 
\renewcommand{\theequation}{A\arabic{equation}}
\setcounter{equation}{0}
\section{Numerical Propagation of S}

If $ H $ is given analytically, then the derivatives of $ H $ can also
be determined analytically.  Therefore, it can be seen from Eqs. (39)
and (40) that the numerical propagation of $ S $ only requires the
numerical evaluation of the partial derivatives of $ S $.  This is
done using centered differences.
  
The $ \q $-grid is given by $ \{(n_1 L_1/N , \dots, n_D L_D/N)| n_i = 0, 
\dots, N-1\} $, giving $ N^D $ grid points.  Note that the $ q_i $'s do not 
need to go all the way to $ L_i $, since $ S(\q + {\bf n}{\bf L}, \bp; t) = 
\bp \cdot {\bf n}{\bf L} + S(\q, \bp; t) $ throughout the evolution.  We track
all $ \bp $ on a grid of canonical momenta about some central momentum
$ \bp_0 $, where our grid consists of all canonical momenta $ \bp_{\bf
k} = \bp_0 + DX {\bf k} $, with $ {\bf k} = (k_1, \dots, k_D) $
satisfying $ |k_1| + \dots + |k_D| \leq GDSZ $.  Let us denote this
set by $ \Omega(\bp_0, GDSZ) $.  Since our evolution involves a
first-derivative in $ \bp $ of $ S $, we cannot compute $
\frac{\partial S}{\partial t} $ at the boundary of the $ \bp $-grid.
The result is that we can only propagate on $ \Omega(\bp_0, GDSZ-1) $,
so that at each iteration the value of $ GDSZ $ shrinks by $ 1 $.
This collapsing boundary method is described in further detail in
Ref. 1, since it also arises naturally in the numerical implementation
of the $ H $ evolution.

Once $ \frac{\partial S}{\partial t} $ has been evaluated on all
possible grid points, the propagation by some time step $ DT $ is done
using the Explicit Euler method, which means that we set $ S(\q,
\bp_k; t+DT) = S(\q, \bp_k; t) + DT \frac{\partial S}{\partial t}(\q,
\bp_k; t) $.

Finally, suppose we are considering a system with unbound degrees of
freedom, that is, some of the $ L_i = 0 $.  Then we track those $ q_i
\in \{q_{i0} \pm n \Delta| n = 0, \dots, N_i \} $.  At each time step,
we can only compute $ \frac{\partial S}{\partial t} $ up to $ n = N_i
- 1 $, so that after each time step we shrink our set of $ q_i $ by
decreasing $ N_i $ by one.

\renewcommand{\theequation}{B\arabic{equation}}
\setcounter{equation}{0}
\section{An Additional First-Order Result}

In this section we derive the first-order result for $ H(\q, \frac{\partial S}
{\partial \q}; 0) $.  Following the procedure in Section 4, we write $ S =
\q \cdot \bp + G(\q, \bp; t) $.  We also write $ H(\q, \p; 0) = H^{(0)}(\p; 0)
+ H^{(1)}(\q, \p; 0) $.  Using $ \p = \frac{\partial S}{\partial \q} =
\bp + \gdq G(\q, \bp; t) $, we get to first-order that,
\begin{equation}
H(\q, \p; 0) = H^{(0)}(\bp; 0) + \gdP H^{(0)}(\bp; 0) \cdot \gdq G(\q, \bp; t)
+ H^{(1)}(\q, \bp; 0)
\end{equation}
For simplicity, we assume that $ G $ was chosen using $ H^{(0)}(\bp) =
H^{(0)} (\bp; 0) $.  As mentioned before, in the first-order limit all
three prescriptions for choosing $ G $ are equivalent.  Using the
other two prescriptions will lead to at most second-order corrections
in our final result.  From Eq. (47) we get that,
\begin{eqnarray}
\gdP H^{(0)} \cdot \gdq G(\q, \bp; t) & = &
-\frac{1}{V}\sum_{{\bf k} \neq 0}{H^{(1)}_{\bf k}(1 - e^{-4\pi^2 ({\bf k} 
\cdot \gdP H^{(0)})^2 t}) e^{2\pi i {\bf k} \cdot \q}} \nonumber \\
                                      & = &
-H^{(1)} + \frac{1}{V}\sum_{{\bf k} \neq 0}{H^{(1)}_{\bf k}e^{-4\pi^2
({\bf k} \cdot \gdP H^{(0)})^2 t} e^{2\pi i {\bf k} \cdot \q}}
\end{eqnarray}
and so we obtain,
\begin{equation}
H(\q, \frac{\partial S}{\partial \q}(\q, \bp; t); 0) = H^{(0)}(\bp; 0) +
\frac{1}{V}\sum_{{\bf k} \neq 0}{H^{(1)}_{\bf k}(\bp; 0)e^{-4\pi^2({\bf k} 
\cdot \gdP H^{(0)})^2 t} e^{2\pi i {\bf k} \cdot \q}}
\end{equation}

\renewcommand{\theequation}{C\arabic{equation}}
\setcounter{equation}{0}
\section{Propagation Time}

Recall from Eq. (32) that the first-order expression for the evolution of
$ H $ in Fourier space is,
\begin{equation}
\frac{\partial H_{\bf k}}{\partial t} = 2\pi i ({\bf k} \cdot \gdP H^{(0)})
G_{\bf k}
\end{equation}
This equation was then used to obtain the gradient-descent prescription
for choosing $ G $ in the first-order limit.  For weak perturbations,
this prescription no longer coincides with the gradient-descent approach,
but should still shrink the nonzero Fourier components of $ H $.  This will
occur as long as the right side of Eq. (C1) (or Eq. (32)) is sufficiently 
dominant compared to the remaining terms in the full PDE for $ H $.  Note
then that for resonances and near-resonances this condition does not hold.
However, for sufficiently non-resonant terms this condition does hold.
Thus, in general, for a weak perturbation, our PDE-based approach starts
out by decreasing the more non-resonant terms of $ H $.  The 
$ \Q $-dependence of $ H $ starts decreasing, and so the rate of change
of $ S $ decreases as well as the evolution proceeds.  Eventually,
the sufficiently non-resonant terms of $ H $ are reduced to a point where
higher-order terms become important, so that our first-order gradient-descent
prescription for choosing $ G $ will no longer work to reduce the 
$ \Q $-dependence of $ H $.  The rate of change of $ S $ then begins to
increase after this point, and eventually the PDE becomes numerically unstable.
By tracking $ \sqrt{\langle (\partial S/\partial t)^2 \rangle} $ on the
grid, it is possible to stop the evolution where the rate of change of $ S $
reaches its minimum, and consequently where the canonical basis has been 
optimized.

Of course, the weaker the perturbation, the closer a given $ {\bf k} $
must be to a resonance for our choice of $ G $ to no longer work to
reduce the corresponding $ H_{\bf k} $.  Futhermore, the weaker the
perturbation, the longer it is possible to propagate the PDE before $
\sqrt{\langle (\partial S/\partial t)^2 \rangle} $ reaches its
minimum, and the closer this minimum will correspond to a
steady-state.  It would be interesting to develop a simple criterion
to estimate at what time this minimum occurs, and how far away the
system is from steady-state at the minimum.

\end{document}